\documentclass[sigconf, screen]{acmart}
\usepackage{url}

\renewcommand\footnotetextcopyrightpermission[1]{} 
\begin{document}

\title{Evaluating Saliency Map Explanations for Convolutional Neural Networks: A User Study}

\author{Ahmed Alqaraawi}
\affiliation{
  \institution{UCL Interaction Centre}
  \city{London}
  \country{United Kingdom}}
\email{ahmed.alqaraawi.16@ucl.ac.uk}

\author{Martin Schuessler}
\affiliation{
  \institution{Technische Universit{\"a}t Berlin\\Weizenbaum Institut}
  \city{Berlin}
  \country{Germany}}
\email{schuessler@tu-berlin.de}

\author{Philipp Wei\ss}
\affiliation{
  \institution{Technische Universit{\"a}t Berlin\\Weizenbaum Institut}
  \city{Berlin}
  \country{Germany}}
\email{philipp@itp.tu-berlin.de}

\author{Enrico Costanza}
\affiliation{
  \institution{UCL Interaction Centre}
  \city{London}
  \country{United Kingdom}}
\email{e.costanza@ucl.ac.uk}

\author{Nadia Berthouze}
\affiliation{
  \institution{UCL Interaction Centre}
  \city{London}
  \country{United Kingdom}}
\email{nadia.berthouze@ucl.ac.uk}

\begin{abstract}
Convolutional neural networks (CNNs) offer great machine learning performance over a range of applications, but their operation is hard to interpret, even for experts.
Various explanation algorithms have been proposed to address this issue, yet limited research effort has been reported concerning their user evaluation.
In this paper, we report on an online between-group user study designed to evaluate the performance of ``saliency maps'' - a popular explanation algorithm for image classification applications of CNNs.
Our results indicate that saliency maps produced by the LRP algorithm helped participants to learn about some specific image features the system is sensitive to.
However, the maps seem to provide very limited help for participants to anticipate the network's output for new images.
Drawing on our findings, we highlight implications for design and further research on explainable AI.
In particular, we argue the HCI and AI communities should look beyond instance-level explanations.
\end{abstract}

\maketitle

\section{Introduction}
As Machine Learning (ML) increasingly becomes an integral part of many computer programs, its impact on our society spans a wide spectrum of domains.
\begin{figure}[h]
\centering
\includegraphics[width=\linewidth]{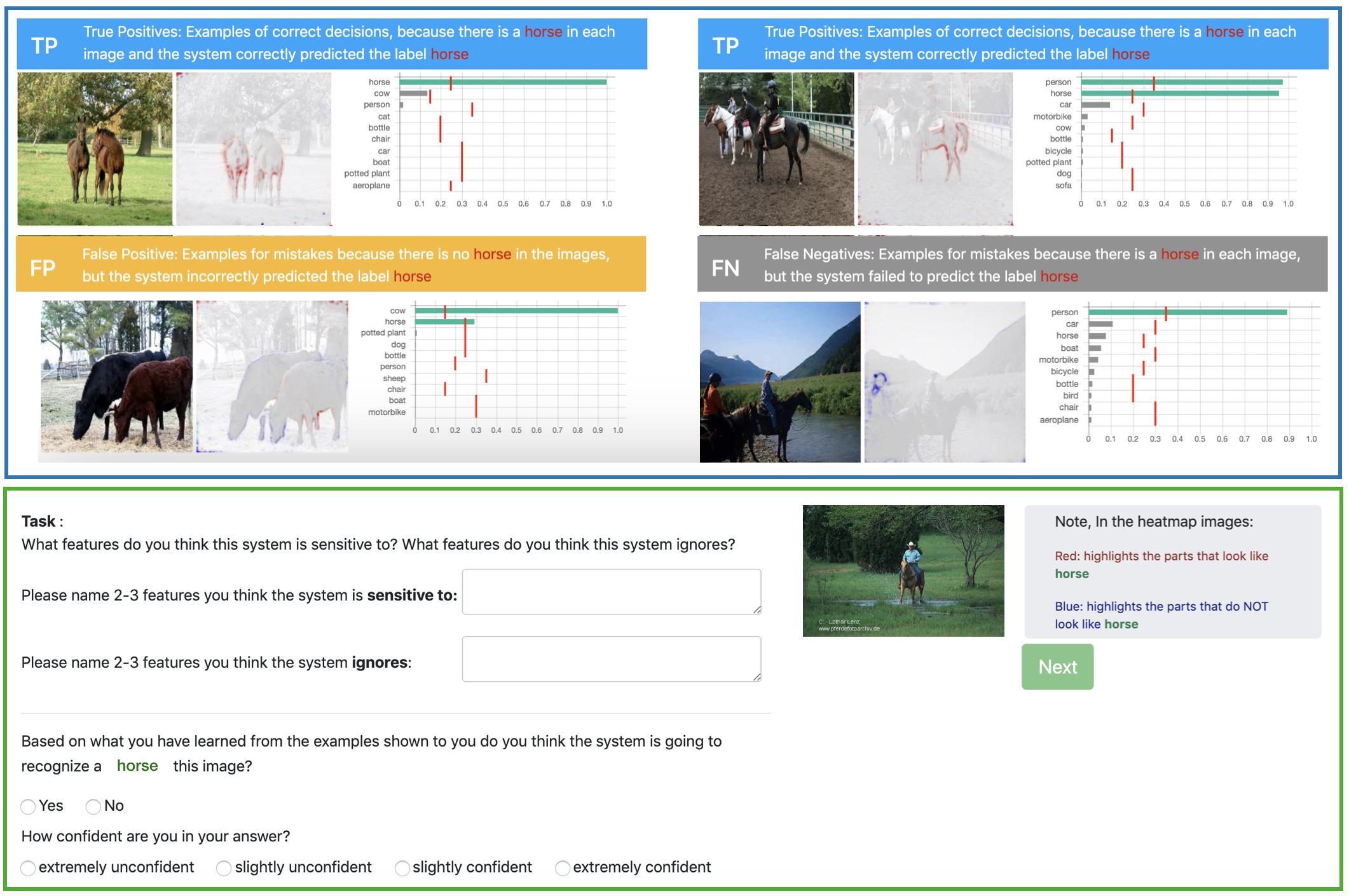}
\caption{The interface: Examples are presented in the blue box at the top. The task is shown in the green box at the bottom. All participants worked on the same tasks and where shown the same examples. Conditions differed only in terms of the additional information that was presented alongside each example. Here, saliency maps and scores are shown.}
\label{fig:interface}
\end{figure}
Some systems have already been shown to outperform humans at certain tasks like lung cancer screening \cite{ardila2019}.
With the ambition to increase efficiency and reduce cost, many public and private organisations are adopting ``data-driven'' ML systems to support or even take decisions around applications that range from predictive policing \cite{mohler2015}, to healthcare \cite{cai2019}, to social services \cite{kleinberg2018}, and many others \cite{stone2016,campolo2018}.
Therefore, there have been several calls to make such systems accountable, so that even users who are not ML experts could decide when to trust their predictions \cite{shneiderman2016,theieeeglobalinitiativeonethicsofautonomousandintelligentsystems2017}.
However, many ML algorithms currently operate as \emph{black boxes}.
When trained with large amounts of data they may perform very well, but understanding the underlying process by which results are achieved is difficult, even for experts. In other words, transparency is still a fundamental and open technical challenge \cite{lipton2018}. This is especially the case for one of the most popular, and best performing types of ML systems: deep neural networks.

We are particularly interested in image classification using convolutional neural networks (CNNs), as this is an area for which some of the most impressive ML results have been reported to date, and with a broad range of applications \cite{pouyanfar2018}. Within this domain, a popular approach to try and make those systems explainable is to produce ``saliency maps'' (also called ``heat-maps'') that highlight which pixels were most important for the image classification algorithm. The claim is that such explanations are easy to interpret by both novice and expert users, and that they can help to detect unexpected behaviour \cite{lapuschkin2019}, and develop appropriate trust towards the system \cite{ribeiro2016}.
Even though several algorithms have been published to produce such saliency maps from CNNs \cite{doshi-velez2017}, very limited research effort regarding their evaluation with actual users has been reported \cite{narayanan2018, yin2019}.

To address this research gap, in this paper we report on an online user study designed to evaluate the performance of saliency maps generated by a state of the art algorithm: layerwise relevance propagation (LRP) \cite{bach2015}. Overall, 64 participants were asked to predict whether an already trained CNN model would recognise an object in an image based on similar examples, and to explain their prediction -- a metric previously proposed to evaluate how \emph{explainable} a system is (the rationale being that if users understand how the system works, they should be able to predict its output \cite{muramatsu2001}).
We used a full-factorial $2x2$ between-group study design.
Consequently, half of the participants were shown saliency maps for the example images, while the other half were not. Moreover, half of them were shown detailed information about the classification scores produced by the CNN.
Our results indicate that when saliency maps were available, participants answered correctly more frequently than when they were absent (60.7\% vs. 55.1\%, p = 0.045). However, the overall performance was generally low even with the presence of saliency maps.

Our data indicates that saliency maps influenced people to notice saliency-maps-features. However, it is unclear if such explanation draw them away from considering other attributes that are usually not highlighted by saliency maps.

Drawing on our findings we highlight several limitations of saliency maps and the resulting implications for design and further research on explainable AI. In particular, we argue that the HCI and AI communities should explore explanation techniques beyond instance-level explanations.

\section{Related Work}
\label{sec:Related Work}

\subsection{Explaining Predictions with Saliency}
While saliency maps have been used in several applications such as the prediction of human eye fixation on images \cite{8315047, 7780434}, in this paper, we focus on the application of saliency maps to explain the behaviour of a CNN model. A large body of literature proposed a variety of different solutions to improve the intelligibility of machine learning models.
For literature reviews, we refer the interested reader to
\cite{lipton2018,guidotti2018,adadi2018}.
One stream in this research field seeks to explain black-box model predictions with post-hoc explanations without uncovering the mechanisms behind them \cite{lipton2018}.
Solutions range from
rendering of prototypical examples \cite{nguyen2016},
textual explanations \cite{hendricks2016},
to displaying examples that are similar to a given input \cite{kim2014,cai2019}.

A particularly popular group of
techniques is feature-attribution \cite{lipton2018}.
For a given input, a relevance score is calculated for each input feature.
Several approaches for calculating the relevance of input features have been proposed.
One estimation method for calculating relevance scores is sensitivity analysis \cite{simonyan2013deep}. For a given sample, some measure of variation (e.g. the gradient) is evaluated.
This way, a relevance score can be assigned to each input variable.
Given a sample, Layer-wise Relevance Propagation (LRP) \cite{bach2015} produces relevance scores by starting at the output of a NN and propagating the output back to the first layer.
The propagation through the network is governed by different propagation rules.
The resulting explanations can be tuned to have different properties with theses rules.

\subsection{Interacting with interpretable Systems}
How users understand systems is a core research interest of the HCI community.
Consequently, a large body of relevant literature exists.
For example, reflections of the potential impact of deep neural networks (DNN) on interpretability date back as early as 1992 \cite{dix1992}.
Other more fundamental theoretical work is centered around the theoretical construct of mental models,
a users' internal representation of a system\cite{norman2014,moray1999}.
If mental models are sufficiently accurate,
they enable an interaction with a system that is more efficient \cite{boulay1999,bayman1984,kieras1984} and more satisfactory \cite{cramer2008,kulesza2012}.
However, when flawed they may cause confusion, misconceptions, dissatisfaction and erroneous interactions
\cite{kulesza2015,tullio2007}.
Similarly, the overestimation of a system's intelligence or capabilities has been shown to impact user interaction negatively \cite{alan2016,kizilcec2016}.
This may lead to over-reliance on a system \cite{lee2004}, less vigilance towards system failures \cite{yang2013}
and unrealistic expectations \cite{yang2013}.
Explanations for better system understanding have been investigated in the context of
information retrieval \cite{koenemann1996},
recommender systems \cite{herlocker2000,cramer2008,kulesza2012}
and context-aware systems \cite{lim2009,dey2009}.

However, currently, the research streams on Explainable, Accountable and
Intelligible Systems of the AI/ML and HCI community are relatively
isolated \cite{abdul2018}.
Researchers also seem to ignore the large body of work in social sciences, which provides valuable insights into explanations \cite{miller2019}.
We seek to contribute to bridging the gap between the involved disciplines by evaluating the state of the art explanation techniques with highly complex models.

\subsection{XAI User Studies}\label{related-user-studies}
Several users studies have been conducted around explainable machine learning.
For example,
Poursabzi-Sangdeh et al. \cite{poursabzi-sangdeh2018} conducted experiments with 1250 lay-users.
They found that participants performed better at estimating the outcome of their model
if there were fewer input features (2 vs. 8) and feature weights were revealed (transparent condition).
However,
such results cannot be generalised
because predicting the outcome of a linear model is a matter of performing a simple multiplication, which does not reflect the complexity of current machine learning models.
Narayanan et al. \cite{narayanan2018} studied if a specific presentation affects the amount of time required for the user to perform a task.
Bussone et al. \cite{bussone2015} raised the question of when explanations could be considered harmful.
In a study that targeted primary care physicians to diagnose and treat balance disorders, the system showed two versions to the users: A comprehensive version, which provides an explanation that shows inputs associated with the diagnosis and
a second version, which shows fewer details.
Their findings indicate that users who received a rich explanation from the system developed an over-reliance bias, which lead them to accept results from the system despite knowing the possibility of error.
Yin et al. \cite{yin2019} conducted a study examining how lay-users understand the performance metrics of ML models.
Their work investigates various aspects of users' understanding and trust of the model performance on a hold-out set and how that maps to the post-deployment performance.

\subsection{Evaluations of Saliency Map for text based classifiers}\label{text-user-studies}
Several studies demonstrated the benefits of techniques
that explains the
importance of individual words
for text-based classifiers.
In a study by Kulesza et al. \cite{kulesza2009}, participants improved the performance of a Naive Bayes e-mail classifier.
In the study, textual explanations and bar charts conveyed the importance of individual words.
In some later work \cite{kulesza2015} the authors introduced their principles for explanatory debugging.
They implemented them in a new version of the e-mail classifier.
In addition to bar charts,
the importance of individual words was now visualised in two other ways:
A word cloud and highlighted words in the e-mail.
Riberio et al. \cite{ribeiro2016} conducted two experiments to evaluate an algorithm to generate saliency maps.
In the first experiment, given two classifiers, participants choose the one they believed would generalize better.
The first classifier provided a better test accuracy but would not generalize well.
The authors concluded that explanations are useful in determining which classifier to trust regardless of the hold-out test accuracy.
In the second experiment,
participants were asked to improve the accuracy of the classifier by removing features that do not seem to generalize.
After multiple rounds, participants were able to enhance the post-deployment accuracy.
However, both studies lacked statistical significant tests and did not have a baseline condition (i.e., No-explanation condition).
Consequently, it remains unknown whether participants would achieve similar performance with no explanation.

Lai and Tan \cite{lai2019} demonstrated a trade-off between performance and human agency by exposing participants to varying levels of machine assistance (of an SVM) while they were identifying deceptive reviews.
They found that explanations without the suggestion of a label slightly improved human performance.
Much higher gains were achieved by showing the predicted labels.
Explicitly suggesting strong machine accuracy further improved performance.
They found that the highlighted words
increased the trust of humans in machine predictions, even when they were randomly chosen.
Feng and Boyd-Graber \cite{feng2019} studied AI-supported question-answering in a trivia game with three types of explanations.
They conducted their study with experts and novices, showing that they trust and use explanations differently.
One of their used explanation techniques highlighted matched words in the question and proposed answer.
They found that this helped participants to decide faster on whether to trust the prediction.
Springer and Whittaker \cite{springer2019} conducted two user studies.
They used a system that predicted the emotional valence of participant's written experiences.
Their explanation technique also used the highlighting of words.
Its perceived performance was initially higher and degraded after users interacted with the system.
Explanations were distracting and caused users to realise the system operated differently than they had initially anticipated.
Most notably, users were disillusioned that the system did note take overall writing context into account.
Instead, predictions were based on simple but accurate lexical weightings.
The explanation techniques in the before-mentioned studies highlight words to raises the users' awareness that they were important.
In that sense, they are comparable to saliency maps.
At the same time, the methods used for determining those features are considerably simpler than saliency-based approaches.
While such methods make controlling factors in a user study easier,
they all utilise machine learning models of lower complexity
which is a significant limitation.
It raises the question of whether findings obtained in these studies will also apply to more complex systems as those used for computer vision such as CNNs.

\subsection{User Studies evaluating Saliency Maps for image classification}
Cai et al. \cite{cai2019},
evaluated two kinds of example-based explanations for a sketch-recognition algorithm: normative explanations and comparative explanations.
Normative explanations led to a better understanding of the system and increased the perceived capability of the system.
Comparative explanations did not always improve perceptions, possibly because they exposed limitations.
While highly relevant for this work, this study did
not evaluate saliency maps.
Riberio et al. \cite{ribeiro2016} also evaluated their algorithm for a simple image classifier.
The authors intentionally trained a biased binary classifier that distinguished between wolves and huskies.
Images of wolves had snow in the background, whereas images of huskies did not.
The classifier was therefore biased towards snow.
In the within-subject study,
participants were first shown ten incorrect predictions and asked whether or not they would trust the model.
Secondly, explanations for the same predictions were shown.
The explanations decreased participants trust in the model as intended.
This study used a very simple scenario where a simple binary classier had an obvious bias.
Again it is unclear whether results would apply to more complex scenarios like multi-class classification with a CNN.

To date, CNNs are becoming the default approach for many computer vision problems \cite{pouyanfar2018}.
While numerous post-hoc explanations for CNNs exist, they are rarely evaluated with users.
To the best of our knowledge, the use of saliency maps has not been evaluated with CNNs or models of comparable complexity.

\section{Method}
\label{method}
We designed a between-group online study to evaluate whether saliency maps can help users understanding of a highly complex CNN used for multi-label image classification.
In the multi-label image classification problem, an image can contain multiple objects.
For example, the assignment of the labels ``\textit{horse, train}'' is considered correct if both, a horse and a train are visible in the image.
We choose this problem because in this context, saliency maps have the potential to highlight specific parts of the image that correspond to one label, as well as parts that correspond to alternative labels.

The study included two independent variables that varied between groups, with a full factorial design.
Both were related to the amount of information shown to participants: \emph{presence of saliency maps} and \emph{presence of classification scores}.

A screenshot of the experimental setup is shown in Figure~\ref{fig:interface}.
In the following sections, we lay out a more elaborate description of the study.
At this point, it is essential to point out that we needed to strike a balance between the number of participants, the duration of the study and the variation of experimental factors.

\subsection{Materials}
\subsubsection{Dataset, CNN Model Architecture and Training}
Various public datasets, algorithms and configuration options exist for the multi-class image classification problem.
We used the PASCAL Visual Object Classes dataset (19714 images), because of its popularity, and its limited number of classes (20).

Additionally, we used the Keras library for Python, starting from an existing Keras model trained on the ImageNet dataset \cite{deng2009}, utilizing the VGG16 architecture \cite{simonyan2014}\footnote{https://keras.io/applications/\#vgg16}.
We then fine-tuned the model on the train-val part of the PASCAL VOC 2012 dataset \cite{pascal-voc-2012}, achieving an Average Precision (AP) score of 0.91 on the training-set and 0.73 on a the validation-set. On a hold-out test-set (the PASCAL VOC 2007 test data \cite{pascal-voc-2007}), the AP was 0.74.
We did not train the model to reach state of the art performance. This was an intentional design choice to understand how explanation techniques could facilitate user understanding about the strengths and limitations of the model.

\subsubsection{Saliency Maps and Scores Generation}
A variety of algorithms have been proposed for generating saliency maps.
In our pilot studies, we investigated two popular implementations: LIME \cite{ribeiro2016} and LRP \cite{bach2015}.
With LRP, saliency maps are not restricted to super-pixel patches but highlight contours of objects, which was preferred by most of our pilot study participants.
For this reason and to simplify our setting, we chose to focus on the LRP algorithm only.
Concretly, we used the $\alpha$-$\beta$ propagation rule \cite{bach2015} with $\alpha = 2$ and $\beta = 1$.

\begin{figure}[h]
\centering
\includegraphics[width=\linewidth]{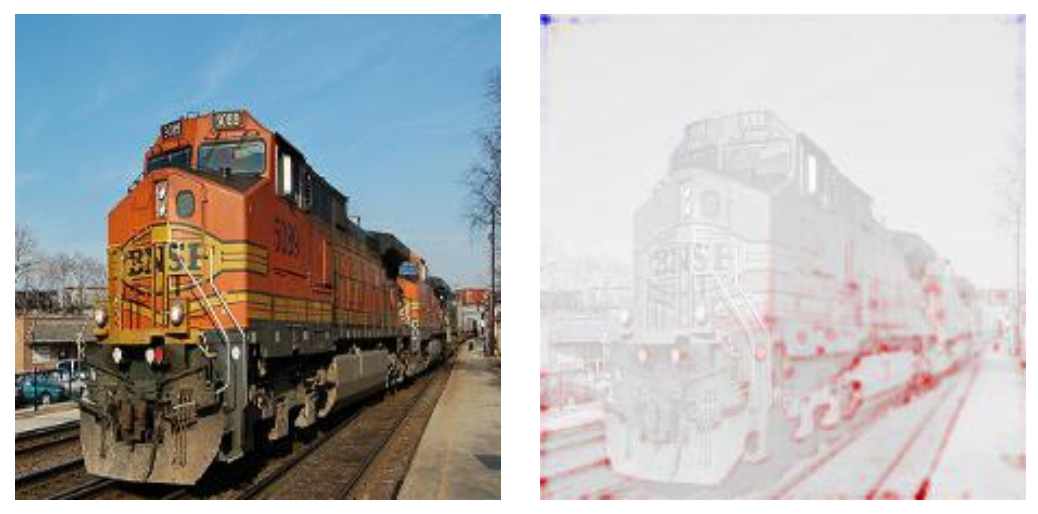}
\caption{Example of a saliency map explanation of a True Positive (TP) image for the label \textit{``train''}. It highlights the contours of the lines below the train. A possible interpretation is that the CNN has learned to recognise trains when rails are present.}
\label{fig:TP_heatmap}
\end{figure}

\begin{figure}[h]
\centering
\includegraphics[width=\linewidth]{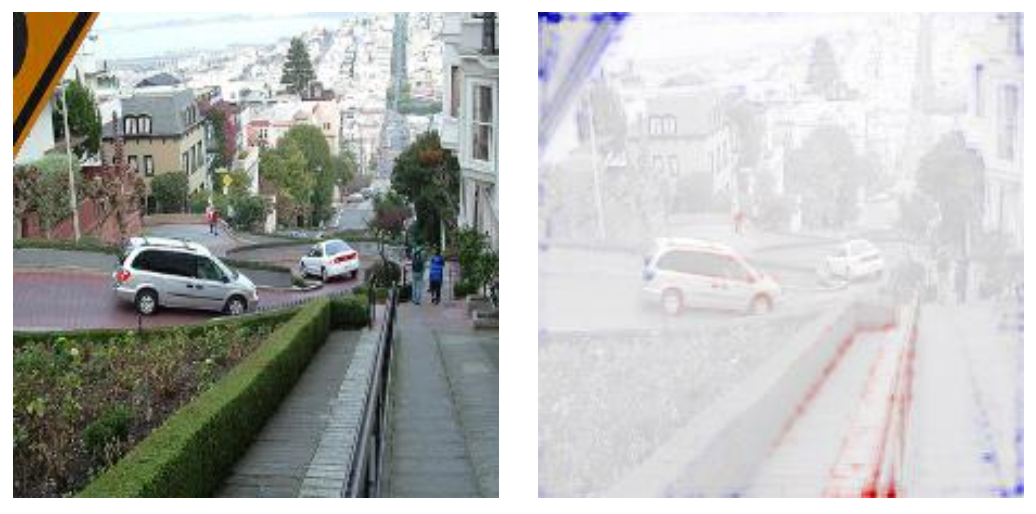}
\caption{Example of a saliency map explanation of a False Positive (FP) image for the label \textit{``train''}. A possible interpretation is that edges in the lower part appeared similar to rails, which could explain this error.}
\label{fig:FP_heatmap}
\end{figure}

Figure~\ref{fig:TP_heatmap} shows a \textbf{true positive (TP)} example, where the model correctly predicts a train. The saliency map suggests that the red part of the image containing the rail supports the classification of this image as a train. Figure~\ref{fig:FP_heatmap} shows a \textbf{false positive (FP)} example where the system \textit{falsely} predicts a train.
The red part of the image contains what \textit{looks like} a rail.
They support the classification of this image as a train. The blue parts are against this classification.

Since an image in the PASCAL VOC dataset can contain multiple objects, for each object class, the CNN computes a classification score between 0 and 1. Hence, a threshold needs to be defined so that the score can be translated into an outcome: \emph{detected} when the score is above the threshold, or \emph{missed} otherwise.
We calculated threshold values for each class (e.g. horse, cat) because the CNN performs differently across classes. In particular, we obtained each threshold by maximising the F1-score for the class on the training dataset. In Figure~\ref{fig:interface}, the small vertical red lines represent these selected thresholds.

\subsubsection{Presentation}
The interface of the study (Figure~\ref{fig:interface}) was implemented as a Web application, using HTML5 and Python with the Django framework. We served the application from a standard Web server. The view-port of the participant browser window needed to be at least a 1000px wide and 600px high during the study.

\subsection{Tasks}\label{tasks}
The user's ability to predict the outcome of a ML classifier has been proposed as a measure to assess how transparent or explainable a system is \cite{lipton2018}. It has also been utilised in other studies \cite{muramatsu2001}.
Thus, we gave our participants the task to predict the classification outcome of the CNN described above for a fixed set of 14 task images from the hold-out test set. More specifically, for each task image, we asked them to list 2-3 features they believe the system is sensitive to and 2-3 features the system ignores.
We then asked participants to predict whether the system will recognise an object of interest (`cat' or `horse') in the given task image. We also asked them to rate their confidence in their forecast on a 4-point forced Likert item. Figure \ref{fig:interface} depicts the interface for one task image (with a reduced number of example images).
Half of the participants started with images of \textit{horses}, while the other half, began with images of \textit{cats}.

To increase participants engagement in the study, in addition to an \pounds8 payment for their time, participants received an additional performance-based bonus of \pounds 0.5 for each correct answer as an incentive.

Seven task images were concerned with the class \textit{``cat''} and another seven with the class \textit{``horse''}. For each task image, participants were shown 12 example images from the CNN training set to inform their judgement.  All participants worked on the same task images and were shown the same example images.

\subsubsection{Selection of Example Images}
We selected the example images for every task image from the PASCAL training set, based on their cosine distance from the task image in the embeddings space generated from the penultimate layer of the network. The assumption was that user understanding might benefit from looking at visually similar images. Showing the outcome of the classifier (i.e. TP, FN and FP) for the examples has been found to be important for the utility of explanation techniques \cite{lai2019}. For this reason, we sampled examples of different outcomes for each task image:
\begin{itemize}
\item 6 examples of True Positives (TP), where a label had been correctly assigned;
\item 3 examples of False Negatives (FN), where the CNN had failed to assign the label;
\item 3 examples of False Positives (FP), where the CNN had incorrectly assigned the label.
  \end{itemize}

  We also based our decision, regarding the number of shown examples, on experience from pilot studies.
  We had noticed that if we presented too many examples, participants were likely to only look at a random subset of them. At the same time, if the number was too low, there was a risk that not enough information was made available to participants.
  For this study, we selected 12 as a compromise. We also noticed that the saliency maps of TP examples are more informative than FN and FP. Thus we decided to show more TP than FN or FP examples.

  \subsubsection{Selection of Task Images}
  We intended our study to be no longer than 40 minutes to avoid fatigue effects. This design choice limited the possible number of task images. Consequently, we had to choose between sampling from a variety of classes or sampling from a subset of classes.
  In our pilot studies, participants found predicting model behaviour very confusing when the class in question was continually switching. Furthermore, the more classes they had to reason about the more challenging the tasks became, because they were not able to ``learn'' much about the model's behaviour regarding a specific class.
  We also wanted to capture a variety of cases where the model had given correct as well as incorrect output.
  For these reasons,
  we decided to limit our experiment to two classes but included three TP, two FN and two FP for each class.

  We drew task images randomly from the hold-out test dataset, with the constraint of having a mid-range classification score. In our pilot studies we had found that images with a low classification score (close to the threshold) were almost unpredictable for participants, while images with a high score were easily predictable. Consequently, we chose to sample from the middle, as we expect to see the most performance variation this way.

  \subsection{Conditions}\label{conditions}
  The study included the following two independent variables:

  \subsubsection{Presence of Saliency Maps}
  This factor had two levels: shown or omitted.
  When shown, the saliency map for the relevant class was displayed next to each example image.
  It is important to note that saliency maps were not shown for the task image but only for the examples.

  \subsubsection{Presence of Classification Scores}
  This factor also had two levels: shown or omitted.
  When shown, a bar chart of the top 10 classification scores was displayed next to each example image.
  Classification scores produced by the CNN are the default sources of explanatory information on the instance level.
  Hence, we aimed to investigate whether visualising this additional numerical information would outperform, compliment or interact with the presence of saliency maps.

  The two independent variables were combined in a full factorial design, resulting in the following four conditions:
  \begin{itemize}
\item Saliency maps not shown and scores not shown (Baseline)
\item Saliency maps not shown and \textbf{scores shown}
\item \textbf{Saliency maps shown} and scores not shown
\item \textbf{Saliency maps shown} and \textbf{scores shown}.
  \end{itemize}

  Figure~\ref{fig:interface} illustrates the \textbf{saliency maps shown} and \textbf{scores shown} condition. In other conditions, the interface looked the same, except not showing the saliency maps or scores.

  \section{Results}

  \subsection{Outcome prediction accuracy}
\begin{figure}[h]
  \centering
  \includegraphics[width=0.9\columnwidth]{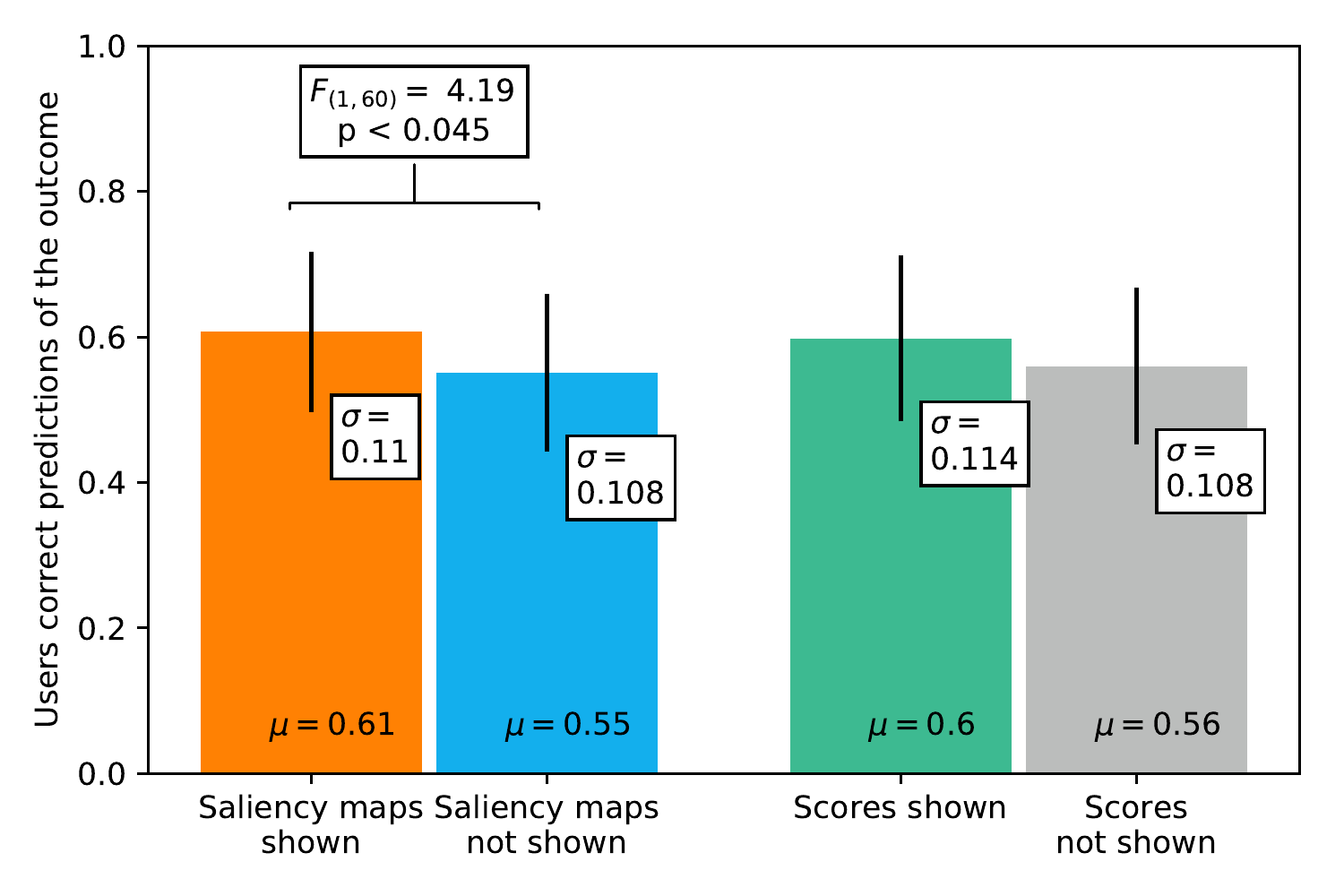}
  \caption{Left: When saliency maps were shown, participants were significantly more accurate in predicting the outcome of the classifier . Right: Scores did not significantly influence the participant's prediction performance.
    Success rates were relatively low across conditions, showing that tasks were very challenging.}
  \label{fig:performance}
  \end{figure}

  We were interested in investigating the effect that the presence of saliency maps and scores has on the ability of participants to forecast the CNN classification outcomes of images.
  We based our performance assessment on the percentage of correct forecasts per participant.
  We summarized the data in Figure \ref{fig:performance}.
  A Shapiro-Wilk test revealed that the percentage of correct forecasts within groups were approximately normally distributed ($W = 0.957, p = 0.027$).
  A Levene's Test showed performance variances between groups were similar ($F_{(3,60)} = 0.156, p = 0.925$).
  \begin{figure*}
  \centering
  \includegraphics[width=1.75\columnwidth]{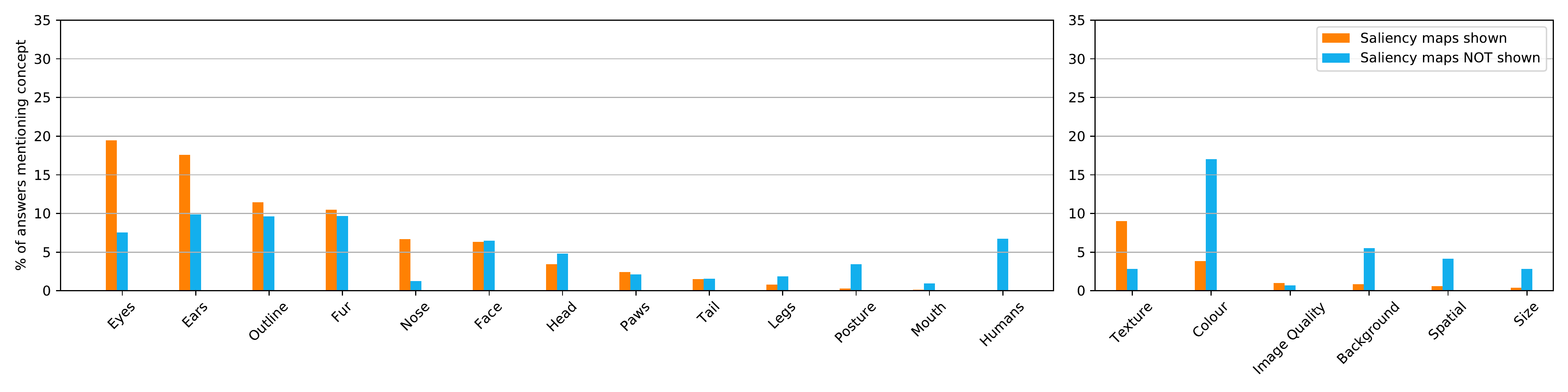}
  \label{fig:freq-cat}
  \end{figure*}
  \begin{figure*}
  \centering
  \includegraphics[width=1.75\columnwidth]{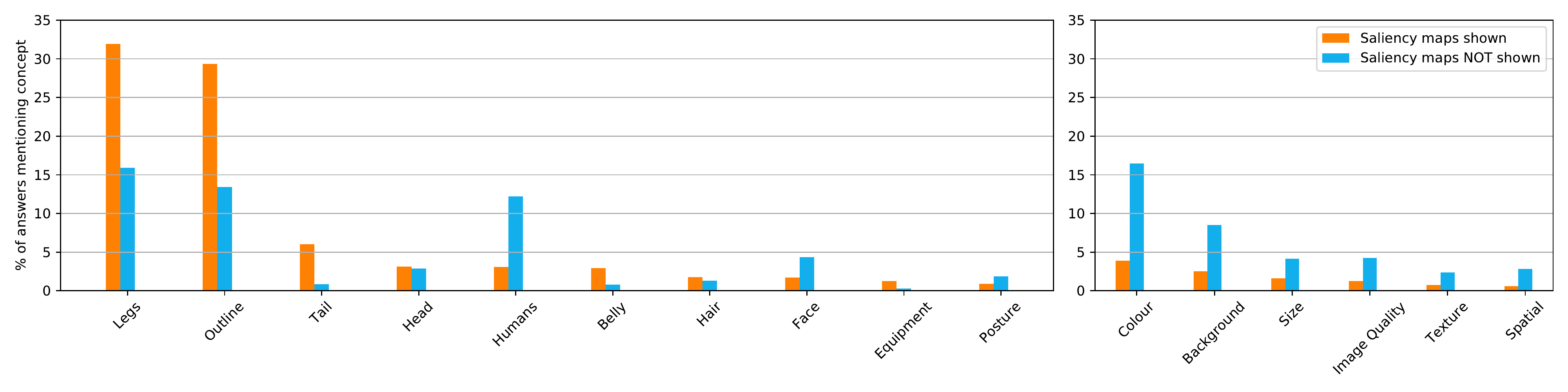}
  \caption{Frequencies of individual features mentioned by participants for images of cats (top) and horses (bottom).  Left: Features belonging to the Saliency-Features. Right: Features belonging to the General-Attributes (frequencies were normalised for each participant).}
  \label{fig:freq-horse}
  \end{figure*}

  A two-way independent ANOVA revealed a statistically significant main effect of the presence of saliency maps on the performance ($F_{(1, 60)} = 4.191, p = 0.045, \eta^2 = 0.063$).
  In the presence of saliency maps participants were more accurate in predicting the outcome of the classifier ($\mu = 60.7\%,  \sigma = 11.0\%$ vs. $\mu = 55.1\%, \sigma = 10.8\%$).
  There was no significant main effect of the presences of scores on performance ($F_{(1, 60)} = 1.938, p = 0.169, \eta^2 = 0.029$).
  Furthermore, there was no interaction effect ($F_{(1, 60)} = 0.060, p = 0.807, \eta^2 = 0.001$).
  \begin{figure}[h]
  \centering
  \includegraphics[width=0.9\columnwidth]{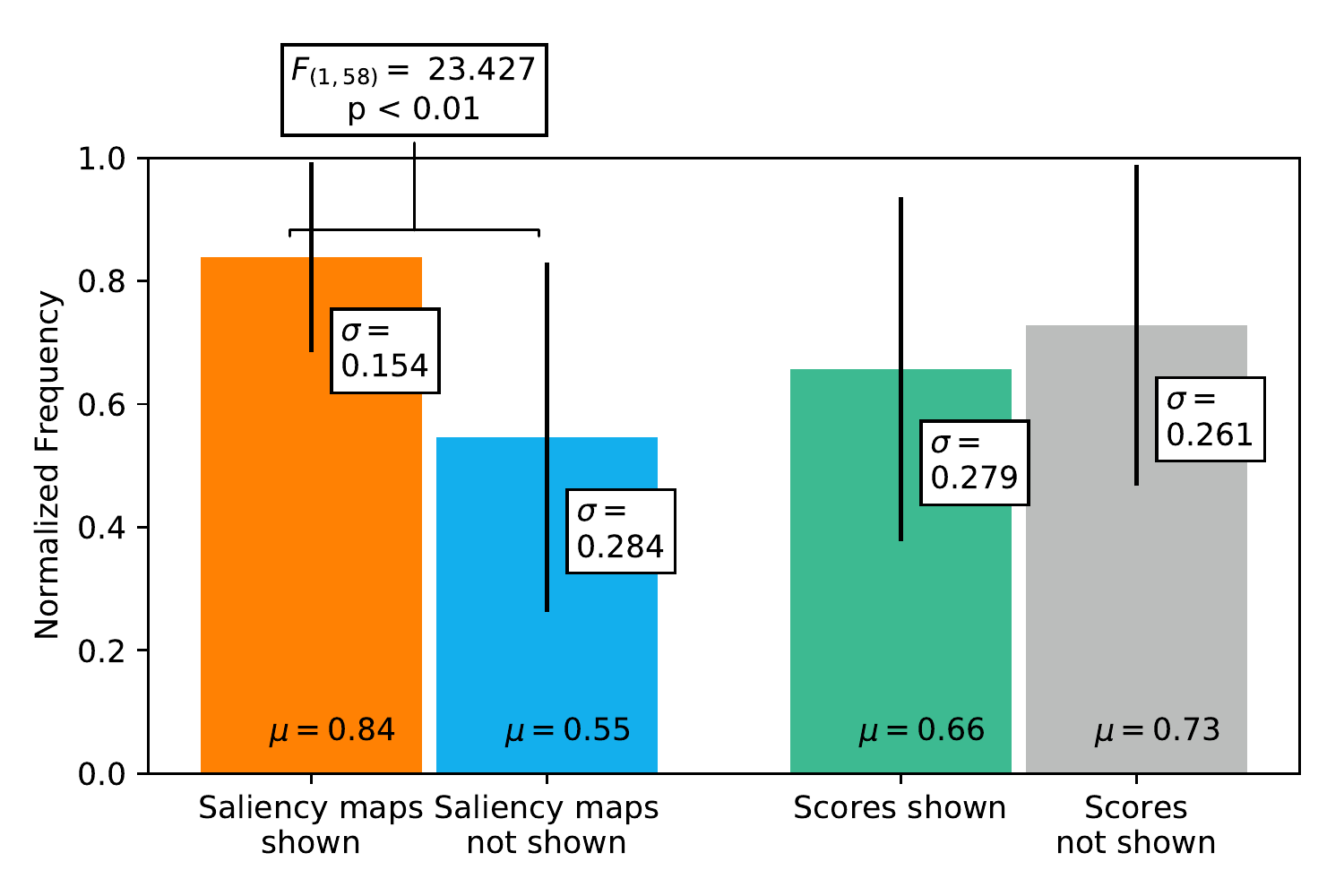}
  \caption{The ratio of mentioned Saliency-Features. It summaries the share of saliency-features participants mentioned per task. They mentioned significantly more such features when saliency maps were present (Left). Scores did not have an influence (Right).}
  \label{fig:features-frequency}
  \end{figure}

  \subsection{Confidence}
  We also asked participants to rate their confidence in their forecast on a 4-point forced Likert item.
  Answers were coded by numbers 1-4 and summed up per participant.
  A one-way independent Kruskal-Wallis test showed that confidence was similar across conditions (H(3) = 1.130, p = 0.770).
  On average participants tended to be \textit{``slightly confident''} in their answers (Median = 3.000).
  We also consider participants' accuracy on the subsets of images corresponding to different outcomes (i.e. TP, FP, FN). Overall the accuracy was higher for TP images, on average 79.4\%, it was lower for FP, on average 46.9\%, and even lower for FN, on average 36.7\%.

  \subsection{Mentioned Saliency Maps Features}
  Besides making a prediction, we asked participants what features they think the classifier is sensitive to and what features it ignored.
  \subsubsection{Excluded data}
  An analysis of the qualitative data revealed that two participants misunderstood these tasks.
  Consequently, they were excluded from this analysis.
  It also became apparent that many of the remaining participants misinterpreted the question about the features the system \textit{ignored}.
  Therefore, we focused only on replies
  participants gave regarding the sensitivity of the classifier to features.

  \subsubsection{Mixed-Method Analysis of Answers}
  We carried carried out a qualitative content analysis \cite{mayring2014}  on the free text replies.
  In the first pass, two of the authors coded the answers inductively.
  Each response could be assigned several open codes based on the features or concepts it addressed.
  Subsequently, coders discussed their individually established codes and agreed on a shared and simplified codebook.
  We decided to assign each code to one of two mayor code groups: \textbf{Saliency-Features} and \textbf{General-Attributes}.

  The \textbf{Saliency-Features} group included codes referring to features, which could be localized to pixels in the proximity of the object of interest and that saliency maps \textit{could} highlight.
  The rationale for this was that we aimed to compare how frequently participants mentioned concepts related features that saliency maps \textit{could potentially} highlight.
  Besides the somewhat obvious feature codes such as \textit{Ears} and \textit{Legs}, this group also included:
  \textit{Equipment} - which applied to all objects associated with domestication such as \textit{``leash''or ``saddle''}, \textit{Outline} which applied to answers referring to the \textit{``shape''} or \textit{``contour''} of the object of interest and \textit{``Fur''} which was used for utterances referring explicitly to the \textit{``fur'', ``skin''} or texture pattern \textit{on} the animal.

  The \textit{General-Attributes} group included codes that refer to utterances of generic properties of the image.
  An example is the code \textit{Background} - which applied to answers referring generically to \textit{``surroundings''} or \textit{``context''} but also objects in the background such as \textit{``trees''}.
  Another example is \textit{Image Quality} which was used for replies addressing issues of \textit{``contrast''}, \textit{``blur''} , \textit{``lighting condition''} or \textit{``occlusion''}.
  The code \textit{Texture} was assigned when answers referred to images \textit{``texture''} generically (i.e. ``Fur patterns'' are considered as a Saliency-Features).

  For the quantitative analysis, we counted the number of Saliency-Features codes and General-Attributes codes.
  We noticed that some participants wrote a lot in the qualitative response and therefore mentioned a lot of features, while others did not.
  To prevent this from skewing the results, we calculated a ratio.
  We obtained the \textbf{Saliency-Features ratio} for each participant by dividing the number of Saliency-Features codes  by the total number of Saliency-Features and General-Attribute codes that we had assigned to their answers.
  Therefore a ratio of $0.6$ means that $60\%$ of the features that a participant mentioned
  were Saliency-Features.
  In the same fashion, we calculated ratios for all codes.
  The top of Figure \ref{fig:freq-horse} shows the ratios for the answers participants gave for images of cats, while the bottom of Figure \ref{fig:freq-horse} shows them for images of horses.

  The Saliency-Features ratio was subjected to a statistical analysis. The data is summarized in Figure~\ref{fig:features-frequency}.
  A Shapiro-Wilk test revealed that the rate of Saliency-Features within groups were approximately normally distributed ($W = 0.900, p < 0.01$).
  A Levene's Test showed that the variances between groups were significantly different ($F_{(3,58)} = 3.749, p = 0.016$).
  To account for heteroscedasticity we ran a two-way independent measures ANOVA using white-corrected coefficient covariance matrix \cite{white1980}.
  It revealed a statistically significant main effect of the presence of saliency maps on the rate of mentioned Saliency-Features ($F_{(1, 58)} = 23.427, p < 0.01, \eta^2 = 0.295$).
  Participants mentioned a larger share of Saliency-Features when saliency maps were present (M = $83.9\%, SD = 15.4\%$ \textit{vs.} $54.6\%, SD = 28.4\%$).
  There was no significant main effect for the presences of scores ($F_{(1, 58)} = 1.384, p = 0.244, \eta^2 = 0.013$) and no interaction effect ($F_{(1, 58)} = 0.004, p = 0.948, \eta^2 = 0.001$).

  \section{Discussion}
  Through a combination of quantitative and qualitative analysis, the results of our study highlight the potential to use saliency maps as an explanatory tool for non-expert AI users, as well as their limitations.
  In the following subsections, we reflect on the key issues and highlight implications for design and further research.

  \subsection{The utility of saliency maps exists, but it is limited}

  Our results show that when saliency maps were shown, participants predicted the outcome of the classifier significantly more accurately. Scores, instead, did not have a statistically significant effect. However, even with the presence of saliency maps, success rates were still relatively low (60.7\%).
  Hence, the task of estimating the system's predictions on a new image remained challenging.
  This is also reflected by our participant's self-reported confidence in their answers, which was not affected by the presence of saliency maps or scores, and was on average still quite low.
  To explain this moderate outcome, we investigated participants’ performance in more detail on subsets of images corresponding to different outcomes. Participants across conditions seemed to be better in predicting the system's outcome when it was correct (\textit{true positives}: 79.4\%). They were mainly struggling with the prediction of errors, performing worse than chance (\textit{false postives}: 46.9\% and \textit{false negatives}: 36.7\%). An interpretation of this result is that participants are possibly inclined to over-estimate the performance of the systems on challenging cases. Such cases are represented by FP and FN images. In fact, in 67.3\% of all cases, participants predicted that the system would be correct, whereas it was only correct in 42.9\% of the cases.
  One of the envisioned applications of explanations is aiding users in building appropriate trust into a system \cite{dzindolet2003,bussone2015}.
  Unexpected and unpredictable failures of a system affect trust more negatively than those that can be understood and anticipated \cite{lee2004,dzindolet2003}.
  Therefore, it is important that users can understand when the system will fail.
  As detecting errors is a claimed utility of instance-level explanations \cite{ribeiro2016,lapuschkin2019}, we suggest that \textbf{future work} should evaluate this empirically in more detail.
  Our study design did not allow to draw conclusions in this regard because we did not fully counterbalance the order of tasks and True Negatives (TN) were not part of the task set.

  \subsubsection{Reasoning on Examples}
  In our study, we based the sampling strategy on the similarity distance between the task image and the training set.
  The rationale behind this was that people might learn more effectively from examples that are similar in appearance to the task image.
  It might help them to reflect upon the \textit{visually similar} images that the system had successfully classified (i.e. TPs) and images the system had  classified incorrectly (i.e FN, FP). We hypothesised that such contrasting reasoning can help users to understand the system’s causes of successes and failures. However,
  when considering the examples presented to participants, we noticed that the usefulness of FN saliency maps is negligible.
  They usually highlight very little evidence (see i.e. the FN example in Figure~\ref{fig:interface}).
  For FN examples, the actual image and the other saliency maps (TP, FN) become the only source of information for understanding why an example has not been recognised by the system.
  This insight suggests that the utility of saliency maps varies according to the classification score. In other words, a saliency map may highlight what supports the prediction of some class, but it will fail to provide counter-factual evidence, namely, the absence of evidence.

  We would like to emphasise that for a human, it is easy to spot and point to the absence of a feature concept, while it is not for a CNN.
  Humans can easily break down an image into meaningful regions (semantics) \cite{fei2007}. In contrast, CNNs look for patterns in a sub-symbolic fashion that lead to an outcome \cite{bishop2006pattern,lipton2018}.
  Because CNNs do not process data in a ‘semantic‘ fashion, other patterns in an image (which may not belong to the concept) can contribute towards a classification outcome in unexpected ways \cite{lapuschkin2019}. \textbf{An implication} for the design is that we need to develop explanation algorithms that bridge the gap between humans and machines by leading the user to understand that the system is not basing its classification decision on higher-level ‘semantics’ of the image.
  Furthermore, we would like to emphasise that choosing representative examples with their corresponding saliency maps, which summarise the behaviour of the system well, is an under-explored topic.
  New approaches for generating saliency maps and for applying them to various machine learning problems are presented (see review \cite{adadi2018}).
  However, very little work exists that investigates for which instances users should examine salience maps.
  Researchers have acknowledged that users can only inspect a limited number of saliency maps \cite{ribeiro2016}, but to the best of our knowledge, only two works explore sampling strategies \cite{ribeiro2016,lapuschkin2019} - none of which where applicable for this work.
  An important implication, then, is that \textbf{further research} needs to characterise the effect of different sampling strategies of saliency map examples on users interpretation of the system operation.

  \subsection{Saliency maps can help participants notice features}
  Our results clearly indicate that saliency maps influenced our participants to notice the highlighted saliency features and to suggest that such features are important for the classification outcome. The ratio of mentioned Saliency-Features (e.g. \textit{legs, outline}) compared to General-Attributes (e.g. \textit{color, image quality}) was significantly higher when saliency maps were present while scores had no influence (Figure~\ref{fig:features-frequency}).

  This effect can be explored in more detail in Figure \ref{fig:freq-horse}. It shows that saliency maps seem to lead people to pay attention to specific parts of the object of interest. For example, Figure \ref{fig:freq-horse} depicts the share of mentioned features for images of horses. It is evident that some features such as \textit{legs}, \textit{outline}, \textit{tail} and \textit{belly} were mentioned much more frequently by participants exposed to saliency maps, while general-attributes such as \textit{background} and \textit{colour} are mentioned more often when the saliency maps are not shown.
  \subsubsection{Facilitating global model understanding by explaining local features}
  It is worth emphasising that even when users notice features, this does not necessarily imply that they will perform better in predicting the outcome of the CNN or reach a global understanding of the model.
  Saliency maps provide only a visualisation of the importance of pixels in a single image.
  Transferring knowledge about potential features to new images, where they are presented in different orientations, scales, forms and perspectives, is very challenging.
  Furthermore,
  it is hard to get a quantifiable measure of the importance of individual features in an image. Again complexity increases if one attempts to quantify the importance of a feature on new images. In other words, it is difficult to estimate how the classification score would change if a feature would be absent. Would the score go down by a factor of 0.1, 0.2 or 0.6? Moreover, does the presence of different features cause an interaction effect? It is challenging for users to reason about this, especially when considering that CNNs process the input data in a non-linear fashion \cite{bishop2006pattern}.

  \textbf{An implication} for the design of explanation systems, then, is that saliency maps should be complemented by a global measure that explains how sensitive the presence of a feature is to the prediction of some class. For example, how sensitive the presence of \textit{nose} is to the prediction of \textit{cat}? In that regard,
  complementing saliency maps with this additional information could be valuable for users to build quantifiable measures of saliency maps, and perhaps avoid biases that might arise from exploring an unrepresentative subset of the dataset. Kim et al. \cite{kim2017interpretability} proposed an algorithm in that direction, where a user can test how sensitive the model's predictions are to a global concept defined by the user. For example, how important the \textit{strips} concept is to the "zebra" class.

  \subsubsection{The importance of general attributes}
  Another reason why noticing Saliency-features does not necessarily facilitate a better understanding of a model is that general-attributes (e.g. colour, contrast) might influence the classification outcome.
  However, these general-attributes are usually not directly highlighted by saliency maps, because as a more general image property, they can not be localised to individual pixels.
  This points to the previously stated limitation of the expressive capabilities of saliency maps \cite{schuessler2019}.
  In fact, saliency maps might even prime participants to primarily consider only highlighted features, and give less weight to other attributes that are not highlighted but important. In contrast, users preconceptions may cause them to focus on attributes such as the \textit{brightness} of the image, even if it is not a major cause of failure. \textbf{An implication} for design is to develop explanations that convey the right expectation to users. We suggest that saliency maps should be complemented by more global representations of the image features. For example, saliency information could be related to global descriptors of the images, such as overall contrast or brightness measures.

  \section{Limitations}
  The design space for the study we presented was vast.
  Our design choices outlined in Section \ref{method} introduced some limitations, which we make explicit in this section.

  The first limitation is the small number of image classes we considered.
  We decided for this compromise considering the limited time for each session, and the limited knowledge participant would have been able to obtain about class-specific behaviour.
  Future work should run a long-term evaluation (i.e. lasting several days or weeks) to allow participants to explore a large dataset with multiple classes in more depth.

  Another limitation of our design is that we used one specific network architecture (VGG16 \cite{simonyan2014}) and one specific technique to generate saliency maps (LRP \cite{bach2015}).
  With a series of pilot studies, we have tried to identify a combination of both techniques which provided saliency maps that participants found to be informative.
  However, this also means that results might be diffrent with a different combination of techniques.

  A limitation of our analysis is that the study design did not allow us to draw conclusions about users performance for different outcomes types (e.g. TP, FN, FP).
  The reason for this was that we did fully counterbalanced tasks, and True Negatives (TN) were not part of the task set. Future studies should address this limitation and study this aspect in more detail.

  Finally, our participants were required to have a technical background, but we did not control for ML expertise. We see potential to repeat our study with different participant populations, such as ML-experts, or lay users.

  \section{Conclusion and Future Work}
  This paper reported on a between-group user study designed to evaluate the utility of ``saliency maps'' - a popular explanation algorithm for image classification applications of CNNs. We focused on saliency maps generated by the LRP algorithm, for one specific architecture and dataset.
  Our results indicate that saliency maps can help users to learn about some specific image features the system is sensitive to, and enhance their ability to predict the outcome of the network for new images.
  However,
  even with saliency maps present, the CNN model remained largely unpredictable for participants (60.7\% prediction accuracy).
  For misclassified images, prediction accuracy remained  well below chance level (43.8\% for False Negatives and 49.2\% for False Positives ).
  We argue that reaching a solid understanding of how a CNN Model classifies images is not possible with the sole use of instance-level based explanations (of which saliency maps are an example).
  Even with very informative examples, saliency maps can only highlight the importance of features that are localisable to pixel-regions. For these highlighted features, they do not convey a quantifiable measure of their importance for future classifications. At the same time, this focus on regions may divert users attention from other important image properties (such as contrast or lighting conditions).
  We suggest using saliency maps in conjunction with other more global explanation methods.
  Furthermore, we view saliency maps sampling strategies as a promising direction for future research.

  Overall, these findings serve as a reminder that making AI explainable is still very much an open technical challenge, and as AI models become increasingly complex, further studies are necessary to address this topic.  We argue that the HCI community is well placed to contribute to solving this challenge, and we hope that the work presented in this paper can serve as a practical example in terms of study design, and stimulate further HCI interest in this area and collaborations with the AI community.

  \section{Acknowledgements}
  This work was supported by the Engineering and Physical Sciences Research Council A-IoT (EP/N014243/1) project, the German Federal Ministry of Education and Research (BMBF) - NR 16DII113, and by the Center of Excellence in Telecommunication Applications at King Abdulaziz City for Science and Technology, National Science, Technology and Innovation plan (NSTIP).
  The study was approved by the Ethics Committees of UCLIC.
  We want to thank members of the Weizenbaum Institue who helped shape this work: In particular, Fenne gro{\ss}e Deters for her valuable advice on the experimental design.
  Hannes-Vincent Krause helped with questions regarding the quantitative analysis.
  Milagros Miceli and Tianling Yang helped to analyse the example images qualitatively (Results were inconclusive and not reported).
  Esra Eres provided manipulated images during the exploratory phase of this work.
  Furthermore, we would like to thank all the participants of the pilots as well as the online study.
  We are also grateful to the anonymous reviewers for their valuable comments and suggestions to improve the quality of the paper. Data URI: \url{https://doi.org/10.5522/04/11638275.v2}.

  \bibliographystyle{ACM-Reference-Format}
  \bibliography{references}

  \end{document}